# Introducing Vision Transformer for Alzheimer's Disease classification task with 3D input

Zilun Zhang[1], Farzad Khalvati[1,2,3,4,5] for the Alzheimer's Disease Neuroimaging Initiative*

[1]Department of Mechanical and Industrial Engineering, University of Toronto
[2]Department of Medical Imaging, University of Toronto
[3]Institute of Medical Science, University of Toronto
[4]Department of Computer Science, University of Toronto
[5]Department of Diagnostic Imaging, Neurosciences & Mental Health Research Program, The Hospital for Sick Children

**Abstract.** Many high-performance classification models utilize complex CNN-based architectures for Alzheimer's Disease classification. We aim to investigate two relevant questions regarding classification of Alzheimer's Disease using MRI: "Do Vision Transformer-based models perform better than CNN-based models?" and "Is it possible to use a shallow 3D CNN-based model to obtain satisfying results?" To achieve these goals, we propose two models that can take in and process 3D MRI scans: Convolutional Voxel Vision Transformer (CVVT) architecture, and ConvNet3D-4, a shallow 4-block 3D CNN-based model. Our results indicate that the shallow 3D CNN-based models are sufficient to achieve good classification results for Alzheimer's Disease using MRI scans.

**Keywords:** Alzheimer's Disease · Convolutional Neural Network · Vision Transformer · ADNI.

## 1 Introduction

Alzheimer's disease (AD) is a neurodegenerative disease that usually starts slowly and deteriorates progressively. It is the cause of 60–80 % of dementia cases [1]. Memory loss is mild in its early stages, but with late-stage Alzheimer's, individuals lose the ability to respond to the environment [2]. A hallmark of

* Data used in preparation of this article were obtained from the Alzheimer's Disease Neuroimaging Initiative (ADNI) database (adni.loni.usc.edu). As such, the investigators within the ADNI contributed to the design and implementation of ADNI and/or provided data but did not participate in analysis or writing of this report. A complete listing of ADNI investigators can be found at: http://adni.loni.usc.edu/wp-content/uploads/how_to_apply/ADNI_Acknowledgement_List.pdf

Alzheimer's disease is the presence of beta-amyloid, the aggregates of proteins that build up in the spaces between nerve cells [2]. There is no cure for AD, and the only treatment is the removal of amyloid from the brain to reduce cognitive and functional decline [2]. As of 2015, there were approximately 29.8 million people worldwide with AD [3].

In AD context, patients can be categorized to AD, Mild Cognition Impairment (MCI), or Cognitive Normal (CN). Machine learning models can help diagnose AD by classifying a patient into these three categories – AD, MCI, or CN – using data from different modalities (e.g., medical imaging, genomic and clinical data) separately or by combining these modalities together. Machine learning models can also predict whether MCI subjects progress to AD (denoted as pMCI) or remain stable (denoted as sMCI) [4][5].

In this paper, we focus on using medical imaging data, specifically MRI scans, to develop classification models for AD vs. CN. For this task, two major pipelines are widely used in medical imaging. The first one exploits radiomic features from MRI scans and trains a separate classifier to classify if a scan belongs to the AD or CN class. The second pipeline functions to train an end-to-end deep learning model (including a feature extractor and a classifier) to extract deep learning features and then to classify the MRI scans. The former mainly uses predefined, hand-crafted features (e.g., size, shape, intensity, and texture information) with feature reduction methods and machine learning classifiers. Li et al. extract texture and wavelet features from each region of interest of the MRI scans and classify the feature using SVM and Random Forest to get an accuracy of 0.861 and 0.835, respectively [6]. Feng et al. use radiomic feature of amygdala with multivariable logistic regression to obtain an accuracy of 0.90 [7]. Li et al. apply wavelet bandpass filtering and select texture radiomic features to calculate high order radiomic features. Then, a SVM classifier is used with the classification accuracy of 0.915 [8]. Du et al. use radiomic features from segmented Hippocampus MRI with SVM to classify different types of AD with accuracy of 0.77 [9]. All of them are subject-wise models while combining ROI techniques. A shortcoming for radiomics models is that they can only look after a small number of pre-defined features. Since 2012, Convolution Neural Networks (CNNs) models with deep learning techniques have been widely applied to different applications in medical imaging and many of them can achieve higher accuracy and precision than models using radiomic features. CNN models are capable of obtaining useful features without pre-defining them. However, the CNN-based deep learning models tend to have lower interpretability compared to Radiomics and they do not generalize well when trained on small datasets. In addition, the CNN structure is hard to capture non-local information within the image due to the limitation of the size of CNN's receptive field [10].

Most recently, Vision Transformer (ViT) based models [11] have shown great potential on classification [[11]–[13] and detection tasks [14], [15] for commonly used benchmarks such as ImageNet [16], and MSCOCO [17]. However, they have rarely been used in classification tasks related to medical imaging, and particularly AD classification tasks using MRI scans. This may be due to several reasons. First, most ViT-based models require a large amount of data to train (e.g., the JFT300M dataset has 300 million images [18], and ImageNet21K [19] has 12 million images) because these models lack inductive bias compared to CNN-based models Inductive bias is a set of assumptions that the learner uses to predict outputs for given inputs. For example, in CNN inductive bias could consist of local visual structures (such as edges and corners), scale-invariance, and translation-invariance. Moreover, due to the large model size of ViT, the training process becomes unstable and hard to converge. Second, most ViT-based models are designed for 2D images. However, MRI scans are 3D images, and it is difficult to apply a ViT-based model directly to MRI scans.

There were several reasons to hypothesize that ViT might perform better on MRI scans, mainly the good representations of the features of AD should not just focus on local regions of the MRI scan. On the contrary, representations that are fused from different regions of an MRI scan may provide the additional information about associations between these regions, which make them become more important for characterization of AD. ViT introduces a sophisticated framework for organizing and utilizing the attention module to extract non-local features. Traditional CNN-based models have difficulty capturing global relationships within an image due to the constraint of the size of the receptive field with limited model depth. In contrast, the attention map calculated from the operating query, key and value pairs can provide non-local information to the ViT model, resulting in better features. For CNN-based models, the problem of the size of the receptive field is more severe for 3D images (MRI scans) because the number of parameters in the 3D-CNN model are much more than in the 2D-CNN model with the same depth. If the model size is fixed, the 3D-CNN model will have a much smaller receptive field which leads to a degeneration of the model capacity.

The ViT model seems promising in the classification of AD, but it was not originally designed for 3D images. To this end, we implemented a 3D version of ViT that can use 3D MRI scans as input and extract non-local features within the Transformer Encoder. Also, we design a convolutional patch embedding module to make this 3D ViT integrate the advantages of CNN. To compare with ViT-based model, we developed a very shallow 3D-CNN model with a Leaky ReLU activation layer and various normalization layers. We evaluated the performance of the models using the public ADNI [20] dataset, which is summarized in section 3.1.

This paper makes several contributions. First, we designed a ViT-based model, the Convolutional Voxel Vision Transformer (CVVT), which takes 3D images (MRI scans) and theoretically exploits the internal associations between different regions of the image. Although the idea of this Convolutional Voxel Vision Transformer is exciting, we found this model does not generalize well when trained on insufficient amounts of data, resulting in a moderate performance in the test set of ADNI. Second, we proposed a shallow 4-block 3D-CNN model, ConvNet3D-4 with Leaky ReLU activation layer and Instance Normalization which makes the model suitable for limited data and small batch size. The ConvNet3D-4 model achieves the state-of-the-art result in the AD vs. CN classification task.

## 2 Related Work

### 2.1 CNN models for Alzheimer's Disease Classification

Most classification architectures using CNN models for AD diagnosis can be divided into three types: 2D slice-level CNN, 3D patch-level CNN (often involved with Region of Interest, ROI), and 3D subject-level CNN [5].

2D slice-level CNN models are designed for 2D images. Their inputs are 4D tensors with the shape of batch size, number of channels, image width, and image height. CT and MRI scans are 3D images. Many researchers choose to slice 3D tensors to 2D tensors and train the model using these 2D tensors. There are several advantages in doing this: First, more data can be generated by slicing operation; second, mature 2D-CNN models, such as ResNet [21], VGGNet [22], and InceptionNet [23] can be applied; third, 2D-CNN models consume significantly less GPU RAM compared to 3D-CNN models. Valliani et al. [24] first applied ResNet which was pre-trained on ImageNet to the axial slices of MRI scans at subject level with affine data augmentation and concluded that the ResNet structure is better than Shallow ConvNet (accuracy of 0.813 vs. of 0.738). Islam et al. [25] and Qiu et al. [26] investigated the slice selection towards the classification result (accuracy of 0.833).

3D-CNN models have an irreplaceable advantage -- they exploit the spatial relationship of voxels explicitly, which is crucial for isotropic scans. 3D-CNN models are designed for 3D images (tensor) that have the 5D shape of batch size, number of channels, image width, image height, and image depth. The input for 3D patch-level CNN consists of a set of 3D patches extracted from an image (scan). Cheng et al. [27] chose 27 large patches with size of 50×41×40 voxels in each MRI scan and trained 27 CNN models. Then they ensembled the features to make a prediction at subject level (accuracy of 0.872). Mingxia et al. [28] used a smaller patch size (19×19×19 voxels) based on anatomical landmarks (accuracy of 0.911). The advantage of patch-based models is that leveraging information from patches requires less GPU memory consumption compared to models using the entire 3D images. The disadvantage is that the model relies on the architecture design when connecting all patches. However, most of the patches used in these models are not informative because they contain parts of the brain that may not be affected by the disease. Therefore, the region of interest (ROI) based CNN model has been developed. Example for these models include 2D+$\epsilon$ CNN [29] (28x28 patches made of three neighboring 2D slices in the hippocampus, accuracy of 0.828) and 2.5D CNN [30] (concatenation of three 2D slices along the three possible planes: sagittal, coronal, and axial. They used 32x32 patches to train the model with accuracy of 0.888).

As for the 3D subject-level CNN, it takes the entire MRI scan of a subject at once and the classification is performed at the subject level. Korolev et al. [31] (ADNI dataset, with 111 scans, accuracy of 0.80), Senanayake et al. [32] (ADNI dataset, with 322 scans, accuracy of 0.78), and Cheng et al. [33] (ADNI dataset, with 193 scans, accuracy of 0.896) used the entire MRI scans as input to classical deep learning models such as VGG and ResNet. The advantage is that the spatial information is fully integrated. But these models require a large amount of GPU memory and only achieved suboptimal classification accuracy.

## 2.2 Transformer and Vision Transformer

Transformer [34] is a powerful model in the field of Natural Language Processing, which has an encoder-decoder structure. Since Vision Transformer adopts its encoder only [11], we will focus on the structure of Transformer Encoder, which is shown in the right part of Figure 1. The Transformer Encoder takes language tokens (tokenized sentences) with positional encoding as input and exploits the inner relationship of tokens using a stack of pre-defined network structures including the self-attention mechanism (Multi-Head Attention block), the normalization layer (Norm block), and the feed forward network (MLP block) to obtain encoded features.

Unlike the original Transformer which uses language tokens (obtained from sentences) with fixed positional encoding as input, Vision Transformer uses visual tokens (obtained from images) with learnable positional encoding as input. The method for obtaining visual token is presented in the left part of Figure 1. For a given image, the ViT first divides it into patches at first, and flattens each patch to a vector. After applying a linear projection to the patch vectors and concatenating an additional vector of class embedding with the result, a learnable positional embedding is added. The resulted vectors are sent to the Transformer Encoder as visual tokens.

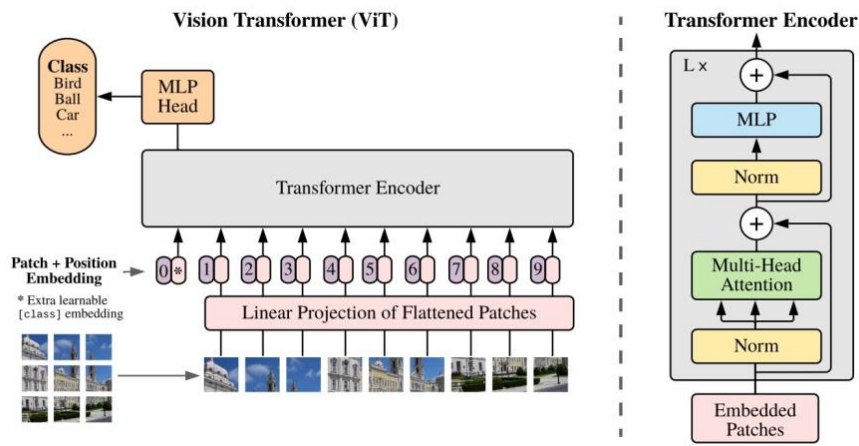

**Fig.1.** Vision Transformer Structure

A key difference between ViT and CNN models is that the ViT uses only attention layers, normalization layers, fully connected layers, and activation functions. Traditional structures used in CNN such as the convolutional layer and the max pooling layer are not used in ViT. This key difference makes the ViT lack the inductive bias (scale invariance, translation-invariance, local visual structures, etc.) and require enormous amount of data to compensate. Another difference is ViT has better scalability and parallel capability than CNN models due to the structure of Transformer. In addition, ViT is good at capturing non-local features within the image compared to CNN models, because the attention map is computed basing on the entire image (patches) rather than the limited receptive field.

ViT achieves impressive results on downstream tasks after fine-tuning. For example, in recognition benchmarks, the best pre-trained ViT model achieves 88.55% accuracy on ImageNet, 94.55% on CIFAR-100, and 77.63% on the VTAB suite of 19 tasks (a benchmark that focus on model transferability) [11]. However, most ViT-based models require a lot of data and resources to train and generalize well. For instance, vanilla ViT uses 300 million images JFT300M and 14 million images ImageNet21K, and the resource consumption varies from 0.23k to 2.5k TPUv3core-days [11]. In medical imaging, Transformer has been introduced to segmentation task (TransUNet) [35]. This model combines the CNN (ResNet50) and Transformer as encoder by feeding the output feature map from CNN as the input sequence of Transformer to extract global contexts. It achieves 77.48 average dice score (74.68 for UNet with ResNet50), and 31.69 average Hausdorff distance (36.87 for UNet with ResNet50) in Synapse multi-organ CT dataset. In addition, a study on replacing CNN with ViT was proposed in [36]. This study performs the classification tasks over three small 2D image datasets, APTOS 2019 (diabetic retinopathy images), ISIC 2019 (dermoscopic of skin lesions), and CBIS-DDSM (mammography images). They use DeiT [13], a ViT variant with distillation technique to compare with ResNet50. The result of ISIC2019 shows that DeiT with random initialization performs much worse than ResNet50 (recall of 0.662 for ResNet50 and 0.579 for DeiT), but the result of DeiT with ImageNet initialization is comparable with ResNet50 (recall of 0.893 for ResNet50 and 0.896 for DeiT). The results in other two datasets point to the similar conclusion. From this paper, ViT-based models have not been widely applied to AD classification because they haven't achieved impressive results in medical imaging classification tasks and hard to train well.

## 2.3   ADNI

The Alzheimer's Disease Neuroimaging Initiative (ADNI) provides researchers with study data. ADNI researchers collect, validate, and utilize data, including MRI and PET images, genetics, cognitive tests, Cerebrospinal Fluid (CSF), and blood biomarkers as predictors of disease. The project began in 2003 [20] as a public-private partnership, led by Principal Investigator Michael W. Weiner, MD.

It has four stages: ADNI-1, ADNI-GO, ADNI-2, ADNI-3 (ongoing). There are three major modalities of data, Imaging Data (3D MRI scans), genetic data (SNPs from genome-wide association study, GWAS), and clinical data (Assessments and Examinations Scores). Many patients' data have all three modalities, along with some basic information such as age, gender, etc.

## 2.4 Leaky ReLU

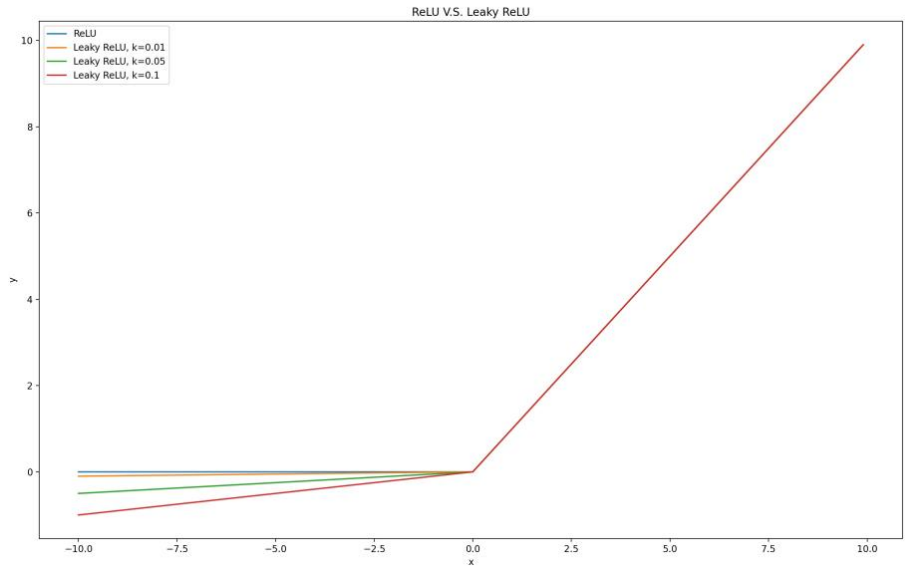

**Fig.2.** ReLU activation function (in blue) and Leaky ReLU activation function with different settings of k (in other colors)

The ReLU (Rectified Linear Unit) [37] activation function is an activation function defined as the positive part of its argument $ReLU(x) = max(0, x)$. Unlike Tanh and Sigmoid activation function, ReLU won't be saturated when input $x$ becomes too large. But when input $x$ is smaller than 0, the input neuron will not be activated since the back propagation value is 0, and that's why Leaky ReLU [38] is proposed. The Leaky ReLU activation function is defined as $LeakyReLU(x) = max(kx, x)$, where $k$ is a hyper-parameter that controls the intensity of the back propagation signal for negative input. The Plot of ReLU and Leaky ReLU with different hyper-parameters is shown in Figure 2.

## 3 Data and Proposed Models

### 3.1 Data

We used T1-weighted MRI data from the ADNI-1 dataset (specifically, the Standardized Image Collections, screening 1.5T), and there are 1075 scans from 818 patients (including patients belonging to MCI class). Some data leakage concerns when using the ADNI dataset are pointed out by [5], such as Wrong data split (data from same patient in different sessions appear in training/validation/test set at the same time), Late split (procedures such as data augmentation or feature selection are applied before isolating test set), Lack of independent test set and Biased transfer learning (test data are used in pretrained model). To avoid these problems, we decided to use the Clinica pipeline to process the dataset.

Clinica [39] is a software platform for clinical neuroscience research studies using multimodal data and most often longitudinal follow-up. Clinica provides a standard data preparation pipeline for ADNI [20], AIBL [40], and OASIS [41] datasets. It organizes and preprocesses the neuroimaging data, then checks the quality and extracts tensors from the pre-processed data.

The Clinica pipeline first converts the NIfTI files (data format of downloaded scans) and ground truth files to the Brain Imaging Data Structure (BIDS data structure) [42]. Then the images are pre-processed with the t1-linear pipeline using the ANTs software package [43] with a sequence of operations including Bias Field Correction [44], Affine Image Registration [45] to MNI152NLin2009cSym Template (Unbiased nonlinear average age-appropriate brain templates in MNI space from birth to adulthood) [46], and Cropping (removing the background, resulting in final images of size 169×208×179×1 (width, height, depth, channel) with 1 $mm^3$ isotropic voxels). The processed data is stored in CAPS data structure [39]. After that, the quality check is performed. Clinica uses a pretrained network that is designed to classify if an image is adequately registered to a template. This procedure is adapted from [47], along with their pretrained models in Github.

When several scans are available for a single subject, the preferred scan, which is listed in MAYOADIRL_MRI_IMAGEQC_12_08_15.csv from the ADNI website is chosen. If a preferred scan is not specified, then the higher quality scan defined in MRIQUALITY.csv is selected. If no quality control is found, then the scan from the first visit is chosen. Images that applied Gradwrap (a system-specific correction of image geometry distortion due to gradient non-linearity), B1-inhomogeneity correction and N3 bias field correction are selected.

The data proportion for each class after the quality check and train-test split is presented in Table 1.

|          | AD  | CN  | Total |
|----------|-----|-----|-------|
| Training | 155 | 189 | 344   |
| Test     | 25  | 25  | 50    |
| Total    | 180 | 214 | 394   |

**Table 1.** Data proportion for each class and train-test split.

## 3.2 Problem Definition

The goal of the classification problem in AD in this paper is to train a model that can perform well in classifying whether a patient had Alzheimer's Disease (AD), or Cognitive Normal (CN) from the MRI scan. Due to the nature of medical imaging for Mild Cognitive Impairment (MCI), it is hard to diagnose MCI only using MRI scan since early MCI shows patterns that are like CN and late MCI shows patterns that are like AD. Other modalities' data such as genomics and clinical data can be used together to help MCI the diagnosis, which is beyond the scope of this paper. We formulate the problem as follows.

For the simplicity in notation, suppose we have N classes and K MRI scans in total. Moreover, we define our classification model $f_{emb} : x \rightarrow v_{emb}$, which takes in an MRI scan $x \in R^{c \times w \times h \times dp}$, and outputs an embedding vector $v_{emb} \in R^d$; a fully-connected layer $fc : v_{emb} \rightarrow v_{logits}$ that maps the feature vector into logit space; and a softmax layer $Softmax : v_{logits} \rightarrow v_{prob}$ that maps the logit into probability space. Note $v_{logits} \in R^N$ and $v_{prob} \in R^N$.

Take an arbitrary training data $x_i \in D^{train}$ that belongs to class $C_k$, and corresponding $y_i = C_k$ ($i \in \{0, 1, ..., K - 1\}$), and all available classes are $C_j$, where $j \in \{0, 1, ..., k, ..., N - 1\}$). The feature vector of this training data in probability space is:

$$v_{prob}{}^i = Softmax(fc(f_{emb}(x_i)))  \quad (1)$$

and j$^{th}$ dimension of the vector $v_{prob}{}^i$ represents the probability of sample $x_i$ belongs to j$^{th}$ class, that is, $P(C_j|x_i)$. We want to optimize $f_{emb}$ and $fc$ so that we can maximize the probability $P(C_k|x_i)$ since $x_i$ belongs to class $C_k$.

### 3.3 Proposed Models

### 3.3.1. Convolutional Voxel Vision Transformer (CVVT)

Papers such as non-local neural networks [10] and Capsule Net [48] claim the spatial and positional information might lose or cannot be used properly by the CNN since the convolution operation is local operation, and the weights are shared. Most CNN-based models only capture local features, and the classifier is trained based on these features. This problem can also occur in the AD classification task. To introduce positional information and attention mechanism, we first developed a ViT-based model (Voxel Vsion Transformer, VViT) with a new patch embedding and positional encoding method. This model uses non-overlapping 3D patches (stride of 50) with a fixed size of 50×50×50 compared to vanilla ViT, which uses 2D patches with a size of 32×32. Finally, 4 · 5 · 4 = 80 3D patches are generated (with padding). A fully connected layer is applied on the flattened patches and positional encoding with the corresponding length is added before sending 3D patches to the Transformer Encoder.

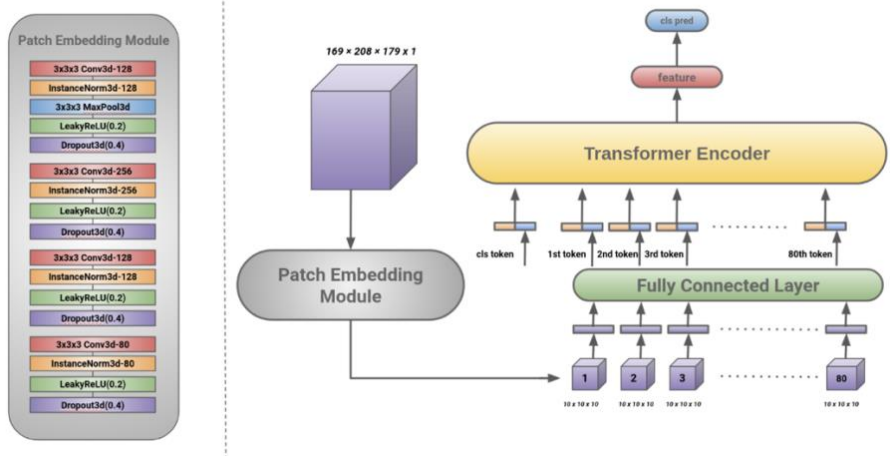

**Fig.3.** Voxel Vision Transformer Model Structure. The purple vectors are flattened 3D patches, the blue vectors are feature vectors after the MLP is applied on flattened 3D patches, and the orange vectors are corresponding positional encodings. Rest details are similar to ViT model.

However, we found this model has extremely poor performance in AD classification tasks. After revisiting the model and calculating the number of parameters of the fully connected layer, we located the problem. Take Transformer-tiny with the token size of 192 as an example, the fully connected layer roughly has 24 M parameters ($50 \cdot 50 \cdot 50 \cdot 192$), but the encoder of Transformer-tiny has only 5.3 M parameters. Imbalanced parameter scales in this scenario will harm the training process by underfitting the fully connected layer and overfitting the encoder of the Transformer.

Thus, we introduce a new convolutional patch embedding module, which consists of a series of convolution blocks with a much smaller fully connected layer in the end, and we call it Convolutional Voxel Vision Transformer (CVVT). Specifically, we use a CNN to downsample the input to a much smaller 3D feature map, which has a size of $10 \cdot 10 \cdot 10$ with 80 channels. Then we use each channel of this 3D feature map as a patch to feed into a fully connected layer to obtain the visual tokens. This design is novel and effective. The size of the patch embedding module decreases from 24 M to 2 M, and it is comparable to the size of the Transformer Encoder (5.3 M). The entire MRI scan is used to let CVVT exploit the inter-association between all voxels. Structure detail is shown in Figure 3.

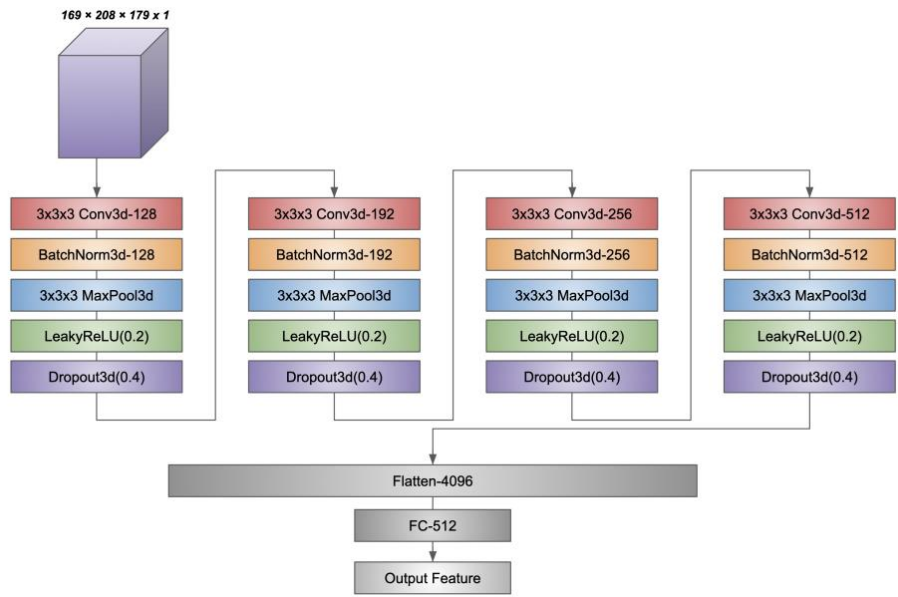

**Fig.4.** ConvNet3D-4 Model Structure

### 3.3.2. ConvNet3D-4

As pointed out in **3.1,** ADNI dataset only has very limited data to train the model. If our ConvNet model becomes too simple like the shallow ConvNet (1-layer) proposed by Valliani et al. [24] the model capacity would be weak. However, if we choose to use complex model (ResNet) like Cheng et al. [33], the model will be hard to train well with limited data.

Therefore, we proposed a shallow yet effective ConvNet3D-4 model, which stacks four Conv-BN-MaxPool-ReLU-DP blocks (Fig 4.) and within each block, the 3D Convolution layer has a 3x3x3 kernel, with stride of 1, padding of 1, and no bias. The kernel size of the 3D Max Pooling layer is 3 as well. The Leaky ReLU [38] is chosen as the activation function and the negative slope is set to be 0.2. The probability of an element to be zeroed in the 3D Dropout layer is 0.4. The channel numbers in each convolution layer across the blocks are $1 \rightarrow 128 \rightarrow 192 \rightarrow 256 \rightarrow 512$. After the last block, the output is flattened, and a fully connected layer is applied in the end to obtain feature with 512-dimension. Since only limited training data (344 scans) is available, this very shallow model is designed to avoid overfitting. Structures such as Conv3d, BatchNorm3d, MaxPool3d, Dropout3d are used in 3D version ConvNet to directly exploit the spatial relationship and feature within the input MRI scan.

## 3.4 Objective Function

The loss function is directly defined by the regular classification loss, Cross Entropy Loss.

$$L_{cls} = L_{CE}(v_{prob}, y)$$

$$= -\frac{1}{K} \sum_{i=0}^{K-1} onehot(y_i) \cdot log(v_{prob}^i)$$

$$where \quad v_{prob}^i = Softmax(fc(f_{emb}(x_i)))$$

(2)

# 4  Experiments

## 4.1  Training Schema

The model is trained using AdamW [49] optimizer, with gradually warm-up

(increasing) learning rate strategy (multiplier=1, total epoch=10) [50]\*and StepLR (linearly decreased learning rate scheduler). We trained the model for 100 epochs using single 1080ti with batch size of 1 due to the constraint from GPU RAM. We set the embedding size $d$ to be 512 ($v_{emb} \in R^d$). The grid search is performed over some hyper-parameters in Table 2.

| Hyper-parameter Name | Choice |
|---|---|
| learning rate | [ 0.01, 0.001, 0.0001] |
| weight decay | [ 0.001, 0] |
| number of epochs decay the learning rate | [25, 40, 80] |
| decay ratio of the learning rate | [ 0.3, 0.5, 0.9] |

**Table 2.**    Hyper-parameter searched.

## 4.2    Evaluation Protocols and Experiment Setup

We trained our model using the training set $D^{train}$, and validated our model in each epoch using the validation set $D^{val}$. All the data were pre-processed with Clinica pipeline mentioned in section 3.1. We trained and evaluated our model as a regular classification task and reported the mean accuracy.

For CVVT model, we select 3 sizes: "tiny", "small", and "base", which are directly adopted from DeiT [13] model, and the detailed settings can be found in Table 3. As mentioned earlier, we had to set the batch size to 1 for the ConvNet3D-4 model. This can lead to unstable batch statistics during training thus harming the model performance if Batch Normalization is used [51]. Therefore, we used the Instance Normalization [52] to replace Batch Normalization. We denote ConvNet3D-4 with Batch Normalization as "ConvNet3D-4-BN", and ConvNet3D-4 with Instance Normalization as "ConvNet3D-4-IN".

| Model Size | Embedding Size | Number of Heads | Depth | MLP Ratio |
|---|---|---|---|---|
| Tiny | 192 | 3 | 12 | 4 |
| Small | 384 | 6 | 12 | 4 |
| Base | 768 | 12 | 12 | 4 |

**Table 3.** Model parameters for different size of CVVT.

---

\* https://github.com/ildoonet/pytorch-gradual-warmup-lr

### 4.3 Experiment Results

We compare the performance of CVVT and ConvNet3D-4 models with several state-of-the-art models in binary classification task on the ADNI dataset. We also include the performance of VViT (CVVT without convolutional patch embedding module) as a reference.

As shown in Table 4, our proposed shallow ConvNet3D-4 models are superior to other existing methods (we set a cutoff of 0.74 in accuracy for other works), and they achieve the state-of-the-art performance at subject level. Particularly, ConvNet3D-4-IN reaches 98% accuracy. That indicates small batch size does influence the model capacity when using subject-level data. Besides, overfitting the classes which have the dominant number of scans is not possible since the test set is class balanced. CVVT models only has moderate results, and a possible explanation is that the training set of ADNI is too small and training a ViT-based model from scratch requires an enormous amount of data to overcome the problem of missing inductive bias.

| Study | Author | Accuracy | Approach |
|---|---|---|---|
| Deep Fusion Pipeline [32] | Senanayake et al., 2018 | 0.76 | 3D subject-level |
| CNNs based multi-modality classification[33] | Cheng and Liu, 2017 | 0.85 | 3D subject-level |
| Combination of multi-model CNN[53] | Li et al., 2017 | 0.88 | 3D subject-level |
| Efficient 3D deep convolutional network [54] | Bäckström et al., 2018 | 0.90 | 3D subject-level |
| A Pilot 2-D+$\epsilon$ Study on ADNI[29] | Aderghal et al., 2017 | 0.84 | ROI-based |
| CNNs Using Cross-Modal Transfer Learning[55] | Aderghal et al., 2018 | 0.90 | ROI-based |
| Multimodal models for detection of AD [4] | Venugopalan et al., 2017 | 0.86 | ROI + 3D patch-level |
| Combination of multi-CNNs[27] | Cheng et al., 2017 | 0.87 | 3D patch-level |
| Multiple cluster dense CNN[56] | Li et al., 2018 | 0.90 | 3D patch-level |
| Hierarchical Fully Convolutional Network [57] | Lian et al., 2018 | 0.90 | 3D patch-level |
| Landmark-based multi-instance learning [28] | Mingxia Liu et al., 2018 | 0.91 | 3D patch-level |
| Deep Multi-Task Multi-Channel Learning[58]] | Mingxia Liu et al., 2018 | 0.91 | 3D patch-level |
| **VViT-tiny** | **Ours, 2022** | 0.72 | 3D patch-level |
| **VViT-small** | **Ours, 2022** | 0.72 | 3D patch-level |
| **VViT-Base** | **Ours, 2022** | 0.74 | 3D patch-level |
| **CVVT-tiny** | **Ours, 2022** | 0.84 | 3D patch-level |
| **CVVT-small** | **Ours, 2022** | 0.86 | 3D patch-level |
| **CVVT-Base** | **Ours, 2022** | 0.84 | 3D patch-level |
| **ConvNet3D-4-BN** | **Ours, 2022** | **0.92** | **3D subject-level** |
| **ConvNet3D-4-IN** | **Ours, 2022** | **0.98** | **3D subject-level** |

**Table 4.** Main Result for AD v.s. CN Classification.

In Figure 5, we select some example images (MRI scans) in axial, sagittal and coronal views from our preprocessed test set for visualization (with model ConvNet3D-4-IN). The left image (labeled as Cognitive Normal) and the middle image (labeled as AD) are correctly classified. The right image (labeled as AD) is a difficult example that our model failed.

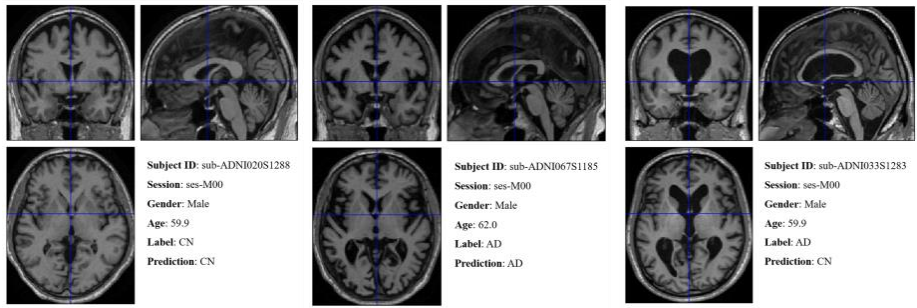

**Fig.5.** Cognitive Normal (correctly classified, left), Alzheimer's Disease (correctly classified, middle), Alzheimer's Disease (incorrectly classified, right)

## 5   Discussion

In this work, we noticed two drawbacks in most existing models. The first one is the small coverage of receptive field in 3D-CNN, and the second one is the lack of nonlocality in traditional CNN-based models. To this end, we designed 3D patch embedding and introduced a purely attention-based model to propose the Convolutional Voxel Vision Transformer (CVVT) with subject-wise 3D MRI scans as input. CVVT is able to spatially explore global and non-local relationships between voxels in an entire MRI scan, which is missing in most existing works on AD classification. However, the classification accuracy of CVVT was lower than most existing CNN-based models, but the idea of communicating features from different regions makes them have great potential for future work.

We have some conjectures about the failure of CVVT, and the first reason is due to insufficient data. Our training set includes only 344 MRI scans. It is hard to train a purely attention-based model from scratch with this amount of data from the experience of Vision Transformer (overfitting and the absence of inductive bias). The second reason for the underperformance could be that spatial relationships in 3D MRI scans are difficult to capture with Transformer Encoder, which was initially designed for 2D images without considering spatial relationships. To improve our CVVT model, we believe that the study of designing an authentic 3D ViT (modifying the structure of the Transformer Encoder to take care of spatial relationship) is important. Also, we haven't tried the pretrained ViT-based model which may bring the inductive bias to our model and potentially improve the classification result. Moreover, the data augmentation and self-supervised learning methods for 3D images could be developed and introduced similar to the solution in [59] to compensate for the problem of limited training data.

To solve the problem of insufficient data when performing classification task on AD, we designed a shallow ConvNet, ConvNet3D-4 to prevent overfitting. Meanwhile, we still wanted to make use of the spatial relationships between voxels and tried to reduce the effect of the small receptive field. We developed a shallow 4-block 3D-CNN model which takes the subject-level MRI scan as input to learn the spatial relationships. Then, we added the max pooling layer with large kernel size to expand the receptive field as much as possible. We had to set the extremely small batch size to fit our model into GPU memory. The small batch size will result in unstable batch statistics for batch normalization, harming the model performance in the end. Therefore, we replaced the batch normalization layer with the instance normalization layer for collecting instance-wise statistics during the training. In addition, we used the Leaky ReLU activation function rather than ReLU in our model to overcome this serious problem of dying neurons since non-zero derivative value will be multiplied to the accumulated gradient for the weight update in back propagation process by Leaky ReLU activation function. The techniques above used in our proposed model are not widely adopted in existing CNN-based models for AD classification. However, they are effective in improving the training stability and avoiding dead neural problem, leading to a better optimization position when limited data is available for training. The results show that the ConvNet3D-4 model performs surprisingly well in the binary classification task (AD vs CN), and even achieves the state-of-the-art result (98% accuracy).

## 6 Conclusions

In this work, to perform the AD classification task, we proposed two models namely CVVT and ConvNet3D-4. The architecture of CVVT was inspired by Vision Transformer. We hypothesized the nonlocality provided by the attention mechanism in CVVT may make it usable for 3D input by designing a 3D embedding module. However, the classification accuracy for this model was not ideal (86% accuracy), and a possible explanation might be data insufficiency which leads to overfitting and the lack of inductive bias. The proposed ConvNet3D-4 model is a shallow 4-block 3D-CNN architecture, which makes use of the instance normalization layer and Leaky ReLU activation function to stabilize the training process. This model achieves a 98% subject-level classification accuracy in a class-balanced test set.

## 7    Acknowledgements

Data collection and sharing for this project was funded by the Alzheimer's Disease Neuroimaging Initiative (ADNI) (National Institutes of Health Grant U01 AG024904) and DOD ADNI (Department of Defense award number W81XWH-12-2-0012). ADNI is funded by the National Institute on Aging, the National Institute of Biomedical Imaging and Bioengineering, and through generous contributions from the following: AbbVie, Alzheimer's Association; Alzheimer's Drug Discovery Foundation; Araclon Biotech; BioClinica, Inc.; Biogen; Bristol-Myers Squibb Company; CereSpir, Inc.; Cogstate; Eisai Inc.; Elan Pharmaceuticals, Inc.; Eli Lilly and Company; EuroImmun; F. Hoffmann-La Roche Ltd and its affiliated company Genentech, Inc.; Fujirebio; GE Healthcare; IXICO Ltd.;Janssen Alzheimer Immunotherapy Research & Development, LLC.; Johnson & Johnson Pharmaceutical Research & Development LLC.; Lumosity; Lundbeck; Merck & Co., Inc.;Meso Scale Diagnostics, LLC.; NeuroRx Research; Neurotrack Technologies; Novartis Pharmaceuticals Corporation; Pfizer Inc.; Piramal Imaging; Servier; Takeda Pharmaceutical Company; and Transition Therapeutics. The Canadian Institutes of Health Research is providing funds to support ADNI clinical sites in Canada. Private sector contributions are facilitated by the Foundation for the National Institutes of Health (www.fnih.org). The grantee organization is the Northern California Institute for Research and Education, and the study is coordinated by the Alzheimer's Therapeutic Research Institute at the University of Southern California. ADNI data are disseminated by the Laboratory for Neuro Imaging at the University of Southern California.